\documentclass[%
 aip,
 amsmath,amssymb,
 reprint,%
]{revtex4-1}

\usepackage{graphicx}
\usepackage{dcolumn}
\usepackage{bm}
\usepackage[colorlinks]{hyperref}

\usepackage[utf8]{inputenc}
\usepackage[T1]{fontenc}
\usepackage{mathptmx}
\usepackage{xcolor}
\usepackage[normalem]{ulem}

\newcommand{\vx}{\mathbf{x}}

\begin{document}

\title{Building intuition for binding free energy calculations: bound state definition, restraints, and symmetry.}

\author{E. Duboué-Dijon}
\email{duboue-dijon@ibpc.fr}
\author{J. Hénin}%
 \email{henin@ibpc.fr}
\affiliation{ 
CNRS, Université de Paris, UPR 9080, Laboratoire de Biochimie Théorique, 13 rue Pierre et Marie Curie, 75005, Paris, France
}%
\affiliation{ 
Institut de Biologie Physico-Chimique -- Fondation Edmond de Rothschild, PSL Research
University, Paris, France
}%
\date{\today}

\begin{abstract}
The theory behind computation of absolute binding free energies using explicit-solvent molecular simulations is well-established, yet somewhat complex, with counter-intuitive aspects.
This leads to frequent frustration, common misconceptions, and sometimes, erroneous numerical treatment.
To improve this, we present the main practically relevant segments of the theory with constant reference to physical intuition.
We pinpoint the role of the implicit or explicit definition of the bound state (or the binding site), to make a robust link between an experimental measurement and a computational result.
We clarify the role of symmetry, and discuss cases where symmetry number corrections have been misinterpreted.
In particular, we argue that symmetry corrections as classically presented are a source of confusion, and could be advantageously replaced by restraint free energy contributions.
We establish that contrary to a common intuition, partial or missing sampling of some modes of symmetric bound states does not affect the calculated decoupling free energies.
Finally, we review these questions and pitfalls in the context of a few common practical situations: binding to a symmetric receptor (equivalent binding sites), binding of a symmetric ligand (equivalent poses), and formation of a symmetric complex, in the case of homodimerization.
\end{abstract}

\maketitle


\section{Introduction}

Binding free energy calculations aim to quantify the binding between two interacting chemical species. Here we focus on explicit, formally exact methods, although many high-throughput approximate methods exist.\cite{Kollman2000,Holderbach2020}
In a biological context, absolute binding free energy calculations are increasingly used in a number of applications,\cite{Chipot2007, Mobley2017} for instance the prediction of ligand-enzyme affinities in the search of effective inhibitors,\cite{Mobley2012,Aldeghi2017} or the study of the stability of protein-protein or protein-nucleic acid complexes.\cite{Jiang2002,Suh2019,Siebenmorgen2020,Jakubec2020} 
Such calculations are typically performed to predict the affinity of a ligand to a biomolecule---for instance in the drug discovery community\cite{Song2020,Cournia2020} to replace long or expensive experiments.
They can also be compared with the available experimental data, for instance, to test the quality of a simulation, which can then be used to gain additional atomic-level insight into the binding process or to decompose it into different components.
Hence, it is crucial to make sure that the calculated binding free energies or binding constants are comparable to experimentally determined values, which requires special care in the theoretical definitions as well as in the computational protocol. 

The computation of absolute binding free energies has been the central topic of a number of works beginning in the 1980s, with seminal theoretical contributions\cite{Hermans1986,Roux1996,Gilson1997,Helms1998, Luo2002,Boresch2003, Mihailescu2004,Woo2005,Mobley2006} that laid the statistical physical foundations of the methods, as well as the practical applications to biological systems.\cite{Hermans1986,Miyamoto1993,Roux1996,Deng2006} 
Thanks to the increase in computational power, the democratization of simulation tools\cite{Mey2020} and availability of pedagogical tutorials, \textit{e.g.} as listed in Ref.~\citenum{TutoWeb}, binding free energy calculations are no longer reserved to a small community of experts, but are applied in a wide variety of contexts, outside the groups that specialize in developing the techniques and software.

However, a few key issues are not as consensual as would be expected in a mature field, and the related literature is still scattered, partly contradictory or confusing, and some notions are hardly accessible to non-specialists.
This is in particular the case of the use of symmetry corrections and the effects of partial sampling, which are, as we will see, intimately linked to the use of binding restraints and to the question of the binding site definition. It is difficult for a non-specialist to acquire a good physical intuition of these complex matters, which, in our experience, can lead to uncritical application of ill-understood formulae, sometimes erroneously, and is hampering the wider adoption of absolute binding free energy calculations.

Our goal here is to present a clear version of the theory, with constant reference to physical intuition, enabling practitioners to understand in depth each step of a calculation and properly handle possibly tricky steps such as standard state, restraint, and symmetry corrections. Only a rigorous treatment of these points makes it possible to obtain well-defined standard free energies and report meaningful comparisons with experimental data. We present typical cases to illustrate common problems encountered in practice, and discuss how to properly treat them, hoping that these will serve as more general guidelines applicable to a broad range of cases.

In this work, we limit our discussion and practical examples to classical all-atom simulations with explicit solvent. For implicit solvent approaches, one may refer for instance to Ref.~\citenum{Swanson2004,Genheden2015}. Two families of methods can be used to quantify binding from simulations: spatial and energy-based.
In the first case, simulations are used to compute the statistical distribution of spatial configurations of the receptor and ligand. This may be done in unbiased simulations, or, more commonly, using enhanced sampling techniques and estimating a potential of mean force (PMF)---or more generally a free energy surface. This case is treated in detail in Section~\ref{sec:pmf}. In the second case, the binding free energy is expressed as function of interaction energies between the receptor, ligand, and solvent.
Statistics on these interaction energies are collected in ``alchemical'' simulations where the potential energy is modified so that the simulations sample often unphysical, but mathematically well-defined states.
The most common form of this approach is the double decoupling method, described in Section~\ref{sec:alchemy}.

In a first part, we present a pedagogical review of the two main classes of binding free energy calculation methods, focusing in more details on the double decoupling method that we will use in our examples.
A second theoretical part is devoted to how to properly account for symmetry (both of ligand or binding sites) and partial sampling.
Finally, we analyze three typical test cases that are chosen to cover the range of situations involving a symmetric ligand, receptor, or complex.

\section{Binding site definition and free energy estimators: a pedagogical review}

We consider a binding equilibrium involving three chemical species: a ``receptor'' $R$, a ``ligand'' $L$, and a complex $RL$.
The receptor could be a macromolecule with a binding pocket that completely surrounds a much smaller ligand, or the structures and roles of both species could be similar: the receptor and ligand can even be the very same species in the case of homodimerization (section~\ref{sec:homodimerization}, where we show, however, that the macroscopic binding constant is different from that for heteroassociation).
For simplicity, we use the conventional names receptor and ligand to cover all cases, but the theory does not depend on each of these having any specific property.
Note, however, that differences between them (of size, flexibility, etc.) do have a practical impact on the convergence of numerical quantities from simulations.
We will refer loosely to those species as molecules, even though they might be non-molecular species such as monoatomic ions or noble gases.

The binding equilibrium writes $R + L \rightleftarrows RL$.
Under constant temperature and pressure conditions, the Gibbs free energy is stationary at equilibrium:
\begin{equation}
\Delta G_\mathrm{bind} = \mu_{RL} - (\mu_R + \mu_L ) = 0
\label{eq:equil}
\end{equation}

From now on, we will assume an \textbf{ideal solution}, which is relevant to many chemical and biological applications. For a more general treatment including the non-ideal case, see Ref.~\citenum{Salari2018}.
If the chemical potentials of the solutes exhibit the ideal concentration dependence, the equilibrium condition \ref{eq:equil} writes:
\begin{align}
    \Delta G^\circ_\mathrm{bind} &= -RT \ln\left(\frac{[RL]C^\circ}{[R][L]}\right)
    \label{eq:equil_dilute} \\
    K^\circ_\mathrm{bind} &= e^{-\Delta G^\circ_\mathrm{bind}/RT} = \frac{[RL]C^\circ}{[R][L]}
    \label{eq:K_def}
\end{align}
where $C^\circ$ is the standard concentration, commonly taken to be 1~mol/L.
Equation~\ref{eq:K_def} is the law of mass action in terms of volume concentrations.
The goal of affinity calculations is to estimate $\Delta G^\circ_\mathrm{bind}$, or equivalently, $K^\circ_\mathrm{bind}$.  Note that the related quantity $K_{\text{bind}}$ is often defined without standard state normalization,\cite{Jorgensen1989,Pranata1991, Shoup1982} which then appears in the alternate binding free energy relationship $\Delta G^\circ_{\text{bind}} = -k_B T \text{ln}\left( K_{\text{bind}} \, C^\circ\right)$.

In principle, a molecular dynamics (MD) trajectory of a solution containing $R$ and $L$ molecules, run long enough to show many binding and unbinding events, provides enough information to compute the affinity.
The probability of association in the microscopic simulation system can be related to the macroscopic binding equilibrium.\cite{Jong2011}
In practice, the time scales required to sample the binding equilibrium are often out of reach, especially in the case of strong binding.
For that reason, direct simulation is rarely used for quantitative affinity estimation.
Furthermore, the theoretical treatment necessary to rigorously connect the microscopic and macroscopic statistics is non-trivial,\cite{Jong2011} and becomes more complex still in the presence of long-range interactions between the solutes.\cite{JostLopez2020}
In the end, this apparently intuitive approach raises considerations that can be counter-intuitive.

To overcome the sampling limitation of the direct approach, binding free energies are thus usually calculated using biased simulations. Two main approaches are most frequently adopted in the literature, namely the Potential of Mean Force (Section \ref{sec:pmf}) and the alchemical double decoupling (Section \ref{sec:alchemy}), whose theoretical foundations we will now recall.

\subsection{$\Delta G^\circ_{\text{bind}}$ from the Potential of Mean Force }
\label{sec:pmf}

We consider a case where the association process is described by a one dimensional potential of mean force (PMF) $w(r)$ along a single coordinate $r$---typically the distance between the ligand and receptor molecules.
The PMF $w(r)$ is related to the radial distribution function $g(r)$ through $w(r)=-RT\ln[g(r)]$.
This does not assume spherical symmetry of the site or the complex, but rather incorporates all information about the statistics of binding into $w(r)$.
We also suppose that the binding site can be delineated by a given range $[r_{\text{min}}, r_{\mathrm{max}}]$ of $r$.
In this case, the binding constant can be obtained by integration of the radial distribution function over the binding site (\textit{i.e.} ensemble of ``bound'' configurations), with the well-known relationship:
\begin{equation}
K^\circ_{\text{bind}} = \dfrac{1}{V^\circ}\int_\text{site} 4\pi r^2 g(r) \text{d}r = \dfrac{1}{V^\circ}\int_{\text{site}} e^{-\beta w(r)} 4\pi r^2 \text{d}r
\label{eq:pmf}
\end{equation}

This widely valid equation has been derived many times via thermodynamic arguments\cite{Gilson1997,Jorgensen1989,Pranata1991,Shoup1982,Luo2002,Gallicchio2011}. It is closely related to the dimer-counting approach mentioned above, as explained in Appendix A.

Application of this PMF approach to complex cases, with the use of restraints and the associated corrections, has been amply discussed from a practical and numerical perspective elsewhere.\cite{Deng2009,Doudou2009,Gumbart2013}

This well-established relation immediately raises three important points. First, as is now usually known, the binding free energy is not simply the well depth of the PMF. \cite{Jorgensen1989}
In the limit of non-interacting species (ideal gas), $w(r)=0$ and $K^\circ_\text{bind}$ simply measures the volume of the binding site with respect to the standard volume: $K^\circ_{\text{bind}} =V_{\text{site}}/V^\circ$.
Hence, and quite counter-intuitively, $K^\circ_\text{bind}\neq 1$ and $\Delta G^\circ_{\text{bind}} \neq 0$.
Second, the standard state has to be explicitly accounted for in binding free energy calculations  by dividing the integral by the standard volume. Finally, $K^\circ_{\text{bind}}$ explicitly depends on a ``site'' definition, which we comment on in detail in section~\ref{sec:site}.
 
\subsection{$\Delta G^\circ_{\text{bind}}$ from alchemical double decoupling}
\label{sec:alchemy}

\subsubsection{Thermodynamic cycle.}
The PMF method to calculate binding free energies, though arguably the most intuitive, becomes difficult to handle and converge for complex geometries. Thus, many applications favor the so-called ``double decoupling'' alchemical approach, which is formally equivalent to the PMF calculation, provided all terms are properly estimated and a consistent definition of the binding site is used.\cite{Deng2009,Corey2019}

\begin{figure*}[ht!]
    \centering
    \includegraphics[width=0.8\textwidth]{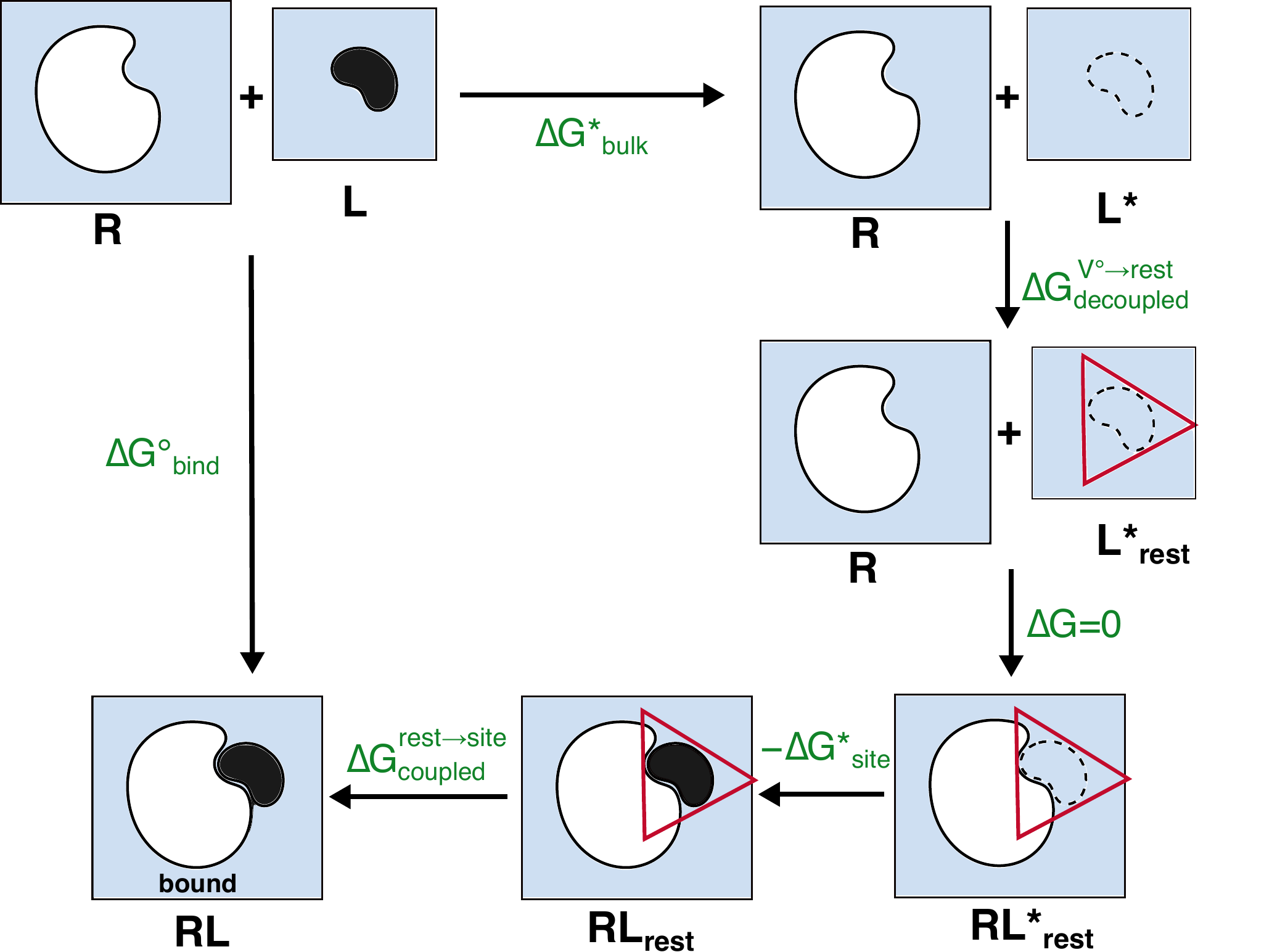}
    \caption{Thermodynamic cycle for double decoupling with ligand restraints. R: unbound receptor. L: ligand coupled to its environment. L*: decoupled ligand. RL: receptor-ligand complex.
    The transformation $R + L^*_\text{rest} \rightarrow RL^*_\text{rest}$ can be seen as a mere translation of the non-interacting ligand, with zero contribution to the free energy.}
    \label{fig:cycle}
\end{figure*}

The Double Decoupling Method builds upon the fact that, since the Gibbs free energy is a state function, $\Delta G^\circ_\text{bind}$ can be calculated using any thermodynamic path between the two states ``RL'' and ``R+L''.
In particular, the double decoupling method\cite{Gilson1997,Pranata1991,Mobley2006,Deng2009} makes use of two alchemical transformations, decoupling the ligand from its environment respectively in the bulk and in the binding site. 
For these respective transformations, free energy differences $\Delta G^*_\text{bulk}$ and $\Delta G^*_\text{site}$ are estimated using numerical estimators such as the exponential formula, Bennett's Acceptance Ratio (BAR)\cite{Bennett1976} or its multi-state variant MBAR,\cite{Shirts2008} the unbinned Weighted Histogram Analysis Method (UWHAM),\cite{Tan2012} or Thermodynamic Integration (TI).\cite{Kirkwood1935}
To improve convergence of these quantities, the transformations go through a number of intermediate states with sufficient overlap.\cite{Pranata1991, Lu2003} If the ligand is charged, the overall charge of the finite simulation box changes during the alchemical transformations, which introduces artifactual free energy contributions that must be accounted for.\cite{Simonson2016,Rocklin2013, Reif2014}

$\Delta G^*_\text{bulk}$ and $\Delta G^*_\text{site}$ are excess free energies of decoupling, and $\Delta\Delta G^*$ is the \textit{excess free energy of binding}, in the sense that in a system of non-interacting particles, their values would be zero.
As a result, the ideal-gas, ``cratic'' contribution to the binding free energy due to translational entropy\cite{Janin1996} can only be accounted for by other terms.
Therefore, the standard binding free energy is not simply obtained as the difference of those two terms---\textit{i.e.} $\Delta G^\circ_\text{bind}\neq \Delta\Delta G^*$ as was sometimes done in the literature\cite{Jorgensen1988,Miyamoto1993,Fujitani2009}---because one has to take into account the standard state reference,\cite{Janin1996} as proposed in 1986 by Jan Hermans and Shankar Subramaniam.\cite{Hermans1986} 

The complete thermodynamic cycle employed is represented in Fig.~\ref{fig:cycle}. Note that modified versions of this cycle can be found in the literature, for instance to facilitate handling ligands with multiple poses.\cite{Sakae2020}
$\Delta G^*_\text{site}$ is the free energy of decoupling the ligand in the ``bound'' state.
In current practice,\cite{Mey2020,Gumbart2013,Salari2018} the corresponding alchemical transformation is performed using a restraining potential $V_{\mathrm{rest}}$ (red triangle in Fig.~\ref{fig:cycle}) that ensures that the ligand stays in the active site during the whole transformation.
In addition to the two decoupling steps already discussed, one needs to evaluate the free energies associated with restraining the decoupled ligand, $\Delta G^{\text{V}^\circ\rightarrow \text{rest}}_\text{decoupled}$---which is often done in the same step as taking into account the standard state reference---and removing the restraints on the coupled ligand in the bound state $\Delta G^{\text{rest}\rightarrow \text{site}}_\text{coupled}$. The desired standard binding free energy $\Delta G^\circ_\text{bind}$ is then obtained by combining all the different steps:\cite{Deng2018}
\begin{equation}
    \Delta G^\circ_\text{bind}=\Delta G^*_\text{bulk}  + \Delta G^{\text{V}^\circ\rightarrow \text{rest}}_\text{decoupled} - \Delta G^*_\text{site} + \Delta G^{\text{rest}\rightarrow \text{site}}_\text{coupled}
    \label{eq:DGcycle}
\end{equation}
Note that in theory, $\Delta G$ values include a $P \Delta V$ term, but that term very nearly cancels out when computing the difference $\Delta\Delta G^* = \Delta G^*_\text{bulk}-\Delta G^*_\mathrm{site}$.
As this term does not intervene in any of the specific issues that we discuss, we will omit it henceforth.

\subsubsection{Use of restraints and corrections.}
\label{sec:restraints}

Restraints not only accelerate convergence of the free energy calculation because they limit the amount of phase space to explore in the (partly) decoupled state(s),\cite{Hermans1986,Helms1995,Gilson1997,Boresch2003,Mey2020,Deng2009,Gallicchio2011,Gumbart2013,Procacci2017} but are also, as previously noted,\cite{Gallicchio2011} necessary from a theoretical perspective, even in the bound state, contrary to what is sometimes stated in the literature.\cite{Fujitani2005,Fujitani2009}
Specifically, restraints are necessary to enforce sampling of (some approximation of) the bound state, because unbinding events, even rare, are incompatible with the estimation of $\Delta G^*_\text{site}$, which is the free energy of decoupling the ligand \textbf{from the binding site} and thus requires the ligand in its coupled state to explore only ``bound'' configurations. Without restraints, the ligand would eventually leave the active site even in the fully coupled state. This is especially apparent in the case of weak binding,\cite{Deng2018} when it happens spontaneously in a relatively short simulation time, but would also be the case even for strong binding in the limit of infinitely long sampling.
Historically, it has been  overlooked because of the kinetic stability of the bound state on the timescale of relatively short simulation times.

To discuss the practical evaluation and magnitude of the two terms, $\Delta G^{\text{V}^\circ\rightarrow \text{rest}}_{\text{decoupled}}$ and $\Delta G^{\text{rest}\rightarrow \text{site}}_{\text{coupled}}$, it is useful to present the commonly used types of restraints. 
While it is tempting to maintain the ligand in the binding site at all stages of the alchemical transformation with a simple restraint on the distance between the receptor and ligand, the free rotation of the ligand allowed by such simple restraints may make sampling of all different conformations complicated, especially in the intermediate stages of the transformation. Explicit restraints on each translational and rotational degree of freedom have thus been suggested,\cite{Mobley2006, Deng2006} possibly in addition to the internal ligand RMSD\cite{Woo2005}, or using an external ligand RMSD as sole restraint coordinate.\cite{Salari2018}
Independently of the chosen degrees of freedom, the restraining potential can be of different forms, typically either harmonic or flat-bottom harmonic, \textit{i.e.} acting only if the chosen degree of freedom is outside a given specified range. Harmonic restraints would restrict the ligand position to remain very close to the reference bound state, while flat-bottom potentials allow for free movement of the ligand in the ``bound'' region.

The free energy associated with adding restraints on the ligand in the decoupled state (\textit{i.e.} gas phase), initially freely evolving in the standard volume, $\Delta G^{\text{V}^\circ\rightarrow\text{rest} }_\text{decoupled}$, does not admit a simple expression in general. With arbitrary complex restraints, one can estimate this term in a separate calculation, for instance gradually switching off the restraints and computing the associated free energy with standard techniques.\cite{Deng2009,Deng2006,Salari2018,Deng2018} For specific forms of restraints however, $\Delta G^{\text{V}^\circ\rightarrow \text{rest}}_\text{decoupled}$  can be obtained more straightforwardly, for instance with an analytical expression when the six body rigid rotations and translations are restrained using harmonic potentials as suggested by reference  \citenum{Boresch2003}. Another simple case is when a restraining potential is applied to the distance between the ligand and receptor. $\Delta G^{\text{V}^\circ\rightarrow \text{rest}}_\text{decoupled}$ can then be numerically estimated by integration of the restraint potential $U_\text{rest}$:\cite{Kumar2015e,DeOliveira2020,Deng2018}
\begin{equation}
    \Delta G^{\text{V}^\circ\rightarrow \text{rest}}_\text{decoupled}= -RT \ln \left(\dfrac{Q}{V^\circ}\right) \;,\text{where}
    \label{eq:DGrest} 
\end{equation}
\begin{equation}
    Q=\int_0^\infty4\pi r^2 e^{-\beta U_\text{rest}(r)} \mathrm{dr} \;.
    \label{eq:DGrest_Q} 
\end{equation}

Finally, one has to estimate the free energy associated with releasing the restraints on the coupled ligand in the binding site $\Delta G^{\text{rest}\rightarrow \text{site}}_{\text{coupled}}$. It is important here to stress that this is \textbf{not} the free energy to remove all restraints on the coupled system and let the ligand freely evolve \textbf{in the whole box}. Instead, it corresponds to the free energy difference between the ligand freely evolving \textbf{in the binding site} and the ligand evolving under the restraints chosen for the alchemical transformation, which are designed to keep it, more or less tightly, in the binding site. Evaluating this term obviously presupposes prior definition of the so-called bound state, \textit{i.e.} the ensemble of configurations that are considered as bound. We will come back to this point in Section~\ref{sec:site}. Evaluation of this term has often been overlooked because restraints are usually chosen to weakly perturb ligand fluctuations in the bound state. In the typical case of strong ligand-protein binding, weak restraints give a negligible contribution in the bound state.
Flat-bottom restraints can be specifically designed so that the unbiased region covers the bound state ensemble, and the restraints therefore have a negligible free energy contribution.\cite{Salari2018} One can even use them directly as the definition of the bound state,\cite{Gallicchio2011} so that by definition $\Delta G^{\text{rest}\rightarrow \text{site}}_{\text{coupled}}=0$. In contrast, harmonic restraints may, depending on their strength, be associated with much more significant corrections. Note that in this framework one can choose as strong restraining potentials as desired for convergence of the alchemical transformation, provided their impact on sampling of the bound state is properly estimated and corrected for.\cite{Wang2006,Deng2006,Deng2009}

In general, if the bound state is defined by an indicator function $I_{\text{site}}(\vx )$, where $\vx$ denotes all the coordinates of the system ($I_{\text{site}}=1$ in the bound state and $I_{\text{site}}=0$ ouside),  then $\Delta G^{\text{rest}\rightarrow \text{site}}_{\text{coupled}}$ can be estimated as:
\begin{equation}
    \Delta G^{\text{rest}\rightarrow \text{site}}_{\text{coupled}} = -RT \ln\left \langle I_{\mathrm{site}}(\vx) e^{+\beta U_{\mathrm{rest}}(\vx)} \right \rangle_{\mathrm{rest}}
    \label{eq:DGrestbound}
\end{equation}
where $\left< \dots \right>$ denotes the ensemble average over configurations in presence of the restraining potential. Eq.~\ref{eq:DGrestbound} can be simply numerically estimated on the fully coupled simulation window if the binding site is well sampled in presence of the restraints (as was for instance done by us and others,\cite{Kumar2015e,DeOliveira2020} where a simplified version of  Eq.~\ref{eq:DGrestbound} was reported, which in those cases is numerically equivalent to Eq.~\ref{eq:DGrestbound}). If this is not the case (for instance when using strong harmonic restraints) then Eq.~\ref{eq:DGrestbound} would not numerically converge and $\Delta G^{\text{rest}\rightarrow \text{site}}_{\text{coupled}}$ has to be calculated through a separate free energy calculation. In particular, if the restraint acts on a single coordinate $z$ and the binding site can also be simply defined by a criterion on that same coordinate, then $\Delta G^{\text{rest}\rightarrow \text{site}}_{\text{coupled}}$ can be calculated from the free energy profile along $z$:\cite{Deng2006}
\begin{equation}
    \text{e}^{-\beta\Delta G^{\text{rest}\rightarrow \text{site}}_\text{coupled}} = \frac{\int e^{-\beta A(z)} I_\text{site}(z)  dz}
    {\int e^{-\beta A(z)} e^{-\beta U_\text{rest}(z)} dz}\\
\end{equation}

\subsection{$\Delta G^\circ_{\text{bind}}$ should be dependent on, but not sensitive to, the site definition.}
\label{sec:site}

From the PMF definition of the binding constant (Eq.~\ref{eq:pmf}), it is clear that $K^\circ_{\text{bind}}$, and equivalently the binding free energy, formally depends on the binding site definition. The integral in Eq.~\ref{eq:pmf} does not converge at large distances, so a range of integration needs to be defined. As discussed above, in the double decoupling framework, a definition of the binding site is also needed to estimate $\Delta G^{\text{rest}\rightarrow \text{site}}_{\text{coupled}}$. This has historically raised many questions\cite{Groot1992,Gilson1997} and is still often a matter of confusion for non-specialists. The origin of the confusion is that it is tempting to argue that since experiments do not need to define the binding site, calculations should not either.\cite{Groot1992} This reasoning, however, has repeatedly been demonstrated to be flawed.\cite{Gilson1997,Luo2002,Mihailescu2004,Gallicchio2011} All theoretical derivations of a binding free energy need to define the configurations that are considered as forming a complex, the bound state, often through a binding site indicator function $I$, taken as 1 for bound configurations and 0 otherwise.\cite{Gilson1997,Gallicchio2011} Alternative expressions using unrestricted integration \cite{Groot1992} do not yield the desired binding free energy but rather relate to the second virial coefficient, probing \textit{non specific interactions} as well as binding.\cite{Mihailescu2004} 
How to define the bound state is then a question that naturally arises. 
As discussed in several works,\cite{Gilson1997,Luo2002,Mihailescu2004} experimental determinations of binding constants implicitly define the bound state as the configurations producing the signal used for detection. Different techniques being sensitive to different types of complexes, the choice of a bound state definition in calculations should in principle depend on the experiment one wants to compare with. For instance, spectroscopic techniques typically detect only closely associated ligand-protein complexes, while calorimetry gives a signal coming from all conformations giving rise to even small heat transfer (even non site-specific association). 
A definition of the bound state from first principles has been proposed for the case where the observable is a phase transition.\cite{Neumann2011}
Explicit calculation of the experimental signal often being out-of-reach, simulations usually assume a two-state model with ``bound'' and ``unbound'' configurations, which is a simplified description of the possibly complex association process. In practice, for strong binding processes, the outcome of the simulations is insensitive to the exact binding site definition, \textit{i.e.} the integration limits in Eq.~\ref{eq:pmf}, so long as all the low free energy regions are included in the bound state. This is illustrated by Fig~1 in Ref.~\citenum{Gilson1997}.
A reasonable choice is to put the boundary at free energy maximum so that the calculated binding constant are relatively insensitive to the its exact location.

\section{Accounting for symmetry}

\subsection{Symmetry number corrections in binding free energies: are they necessary?}

Perhaps the most influential expression for a binding free energy is that proposed by the landmark paper of Gilson et al.\cite{Gilson1997}
In addition to the decoupling free energies and pressure-volume term, their main expression for the absolute binding free energy includes a correction 
\begin{equation}
\Delta G_\text{sym} = RT \ln \left(\frac{\sigma_{RL}}{\sigma_R\sigma_L}\right)\;,
\label{eq:Gilson_sym_correction}
\end{equation}
where $\sigma_x$ denotes the symmetry number of species $x$.
This suggests that the binding free energy should include an explicit term to account for changes in symmetry between the separated components and the complex.

The symmetry term arises from the expressions of the chemical potential of each species based on its classical partition function.\cite{Chandler1976}
When defining a classical partition function, all atoms are labeled as if they were distinguishable.
If the potential energy function treats atoms of the same element identically, the resulting atomic partition function must be divided by the number of permutations of atoms of the same element in the system to compensate for the overcounting.

This is not the case in force-field based classical simulations, where molecules are defined by non-dissociative bonds (usually harmonic) between arbitrarily labeled atoms. Most atom permutations break the bonded structure of the molecule and lead to high energies, so that the resulting configuration does not contribute to the computed partition function. However, if a molecule is symmetric, some permutations of atoms---``symmetry permutations'', for instance, swapping the two oxygen atoms of a carboxylate group---preserve the molecular structure and the energy. 
All symmetry-related configurations thus contribute equally to the partition function.
Therefore, defining non-dissociative bonds leads to double or multiple counting \textit{only for symmetry-related configurations}.
Dividing the partition function by the molecular symmetry number corrects for this overcounting.
To be precise, the relevant molecular symmetry number is the number of permutations of atoms within the molecule that leave the potential energy unchanged, as argued by Gilson and coworkers themselves.\cite{Gilson2010, Gilson2013}
It can also be seen as the symmetry number of the molecular graph, defined by \textit{non-dissociative bonds} in the force field.

If non-covalent binding is described only by dissociative, ``non-bonded'' energy terms, as typically done in force-field simulations of biomolecules, the overcounting of bound ($RL$) and unbound ($R+L$) configurations is precisely the same (the potential energy function has the same symmetry in the bound and unbound states), so that $\sigma_{RL} = \sigma_R \sigma_L$. In this case, the symmetry correction of Eq.~\ref{eq:Gilson_sym_correction} becomes:
\begin{equation}
\Delta G_\text{sym} = RT \ln \left(\frac{\sigma_{RL}}{\sigma_R\sigma_L}\right) = 0\;.
\end{equation}

This runs counter to an intuitive understanding of the symmetry numbers and in particular of the meaning of $\sigma_{RL}$.
Crucially, it is \textit{not} the symmetry number that a chemist would ascribe to the complex; for example, one would ascribe the monodentate acetate-cation complex (Section~\ref{sec:acetate}, Fig.~\ref{fig:acetatesym}) a symmetry number of $\sigma_{RL} = 1$---as opposed to $\sigma_{R} = 2$ for free acetate---because intuitively, the bound cation breaks the symmetry of the carboxylate group.
However, the potential energy remains symmetric with respect to permutation of the two oxygen atoms, regardless of the position of the cation. The correct symmetry number of the complex is thus not 1 but 2, as that of the unbound receptor.
(Rigorously speaking, that number should be multiplied by $3!=6$ to account for permutations of the methyl hydrogen atoms, but that symmetry is of no practical consequence here, and is routinely ignored by practitioners without question.)

The use of often ill-interpreted and counter-intuitive symmetry numbers is a common source of mistakes, including in our own work.\cite{LeBard2012}
The treatment that we propose for double decoupling (Section~\ref{sec:restraints}) eliminates the need for such a posteriori symmetry corrections. By introducing separately a definition of the bound state and a set of binding restraints, all ``symmetry-related'' free energy contributions are naturally accounted for by the two restraint-related free energy terms $\Delta G^{\text{rest}\rightarrow \text{site}}_\text{coupled}$ and $\Delta G^{\text{V}^\circ\rightarrow \text{rest}}_\text{decoupled}$, in the spirit of early work by Hermans and Wang.\cite{Hermans1997}
This seems to us more intuitive, and less error-prone, than relying on abstract symmetry considerations.\cite{Note1}

\subsection{Partial sampling: ``if you've seen one, you've seen 'em all''}
\label{sec:partial_sampling}

In alchemical perturbation simulations, whether or not there are restraints that restrict the symmetry of the bound complex, the ensemble of configurations that are effectively sampled in finite time can be smaller than the true ensemble, and in particular, some modes of a symmetric distribution can be poorly sampled, or not sampled at all.
In this case, it has been argued that unsampled modes lead to missing terms in the calculation of partition functions, which biases the calculated free energies and requires a symmetry correction.\cite{Mobley2006} 
We argue here that this statement is incorrect in the case of an alchemical perturbation.

The argument would be valid only if the partition functions for the relevant states were estimated \textit{separately} by integration over configuration space based on independent simulations.
Then, any region missed in sampling in the complex would be effectively unaccounted for.
However, in alchemical free-energy calculations, this is not the case: partition functions for individual end states are never estimated as \textit{integrals}, but rather, their ratios are computed as ensemble \textit{averages}.
Recall that the free energy of alchemical perturbation from state $A$ to state $B$ can be written\cite{Zwanzig1954}:

\begin{equation}
  e^{-\beta \Delta F_{AB}}  = \left\langle e^{-\beta \Delta U_{AB}} \right\rangle_A
\end{equation}

Numerically this ensemble average is estimated by an average over configurations generated in state $A$.
Suppose that state $A$ exhibits two symmetric modes $\Gamma$ and $\Delta$, from which $n_\Gamma$ and $n_\Delta$ samples, respectively, are collected.
The exponential estimator $\Delta \tilde F_{AB}$ then writes:
\begin{align}
  e^{-\beta \Delta \tilde F_{AB}} &= \frac{1}{n_\Gamma + n_\Delta} \left(\sum_{\vx_i \in \Gamma} e^{-\beta \Delta U_{AB}(\vx_i)}
  + \sum_{\vx_i \in \Delta} e^{-\beta \Delta U_{AB}(\vx_i)}\right) \\
  &= \frac{n_\Gamma}{n_\Gamma + n_\Delta} \left\langle e^{-\beta \Delta U_{AB}} \right\rangle_\Gamma
  + \frac{n_\Delta}{n_\Gamma + n_\Delta} \left\langle e^{-\beta \Delta U_{AB}} \right\rangle_\Delta
\label{eq:partial_sampling}
\end{align}

where the brackets correspond to empirical averages of samples taken from the respective modes. If the modes are symmetric, then configurations from $\Gamma$ and $\Delta$ yield the same distribution of $\Delta U_{AB}$ values, 
and the computed average does not depend on the number of samples drawn from each mode.
Thus, the number of transitions (if any) observed between symmetric modes during the course of the simulation does not impact the convergence nor the accuracy of the free energy estimate.

A significant source of confusion about this question may come from the process of stratification, whereby discrete intermediate states between $A$ and $B$ are simulated to compute the total free energy difference.
This suggests that both end points of the transformation are not sampled identically, and \textit{e.g.} only one of the modes is kinetically accessible in the coupled state, but the whole space is readily sampled in the decoupled state.
This has been argued to require individual symmetry corrections for each ``pair of states'' along the transformation.\cite{Mobley2006}
In a stratified setting, the states $A$ and $B$ above would correspond to intermediate states, not the final end-states.
Then, as shown above, the free energy difference \textit{for each window} is unbiased by unbalanced sampling of symmetric modes.
The fact that this happens to various degrees over different windows is immaterial.
This remains true when combining data from all states at once to estimate the free energy using MBAR\cite{Shirts2008} or UWHAM,\cite{Tan2012}
as these estimators only depend on the statistical distribution of comparison energies between the states, which is itself insensitive to which of the symmetric modes the samples were collected from.

Shortly put, the first step in all free energy estimators is to take a set of configuration samples $\vx$ and map it to a set of comparison energy samples $\Delta U_{AB}(\vx)$ (or for TI, energy derivatives $\partial U(\lambda, \vx)/\partial \lambda$). At this step, symmetry-related configurations map to the same energy, and their distribution among symmetric modes becomes irrelevant. This is illustrated numerically below in the case of acetate-cation binding (\ref{sec:acetate}), where we show that partial sampling of symmetric modes does not affect free energy estimates.

In contrast, Equation~\ref{eq:partial_sampling} gives insight on when partial sampling would result in a biased free energy estimate: this happens if and only if \textit{the energy perturbation $\Delta U_{AB}$ is asymmetric}. That is the case e.g. for restraint free energy perturbation if the restraint potential breaks the symmetry (states $A$ and $B$ have different symmetry). Then, sampling all modes with their correct statistical weight is critical to obtain the correct free energy contribution.

\subsection{The case of symmetry-breaking restraints}
\label{sec:sym_restraints}


A corollary  of the result above, perhaps even more counter-intuitive, is that any restraint that has no other effect than restricting sampling to one of the symmetric modes \textit{does not affect the decoupling free energy}.
Suppose that the same binding process is studied with two numerical approaches, either a symmetry-compatible restraint or a symmetry-breaking restraint.
Despite the difference in restraints, the two approaches give the same values for both decoupling free energies $\Delta G^*_\text{bulk}$ (of course) and  $\Delta G^*_\text{site}$ (because of the arguments above).
One may then wonder if the change in restraint contributions to the free energy causes the two approaches to disagree with each other.
In fact, differences appear in both restraint free energies
$\Delta G^{\text{V}^\circ\rightarrow \text{rest}}_\text{decoupled}$ and
$\Delta G^{\text{rest}\rightarrow \text{site}}_\text{coupled}$, and these two contributions cancel out in Equation~\ref{eq:DGcycle}, yielding the same binding free energy for both schemes.
This is illustrated by a fully analytical treatment of a toy case in Appendix~B.

This is not a purely theoretical consideration, as different applications make one or the other approach more convenient in practice.
For example, complexes with spatially distinct binding sites (\ref{sec:insulin}) are more naturally described with a restraint surrounding a single site.
Conversely, complexes with fast-exchanging binding modes such as the different orientations of a benzene ligand might be easier to simulate with a simple translational restraint encompassing all binding modes.
Therefore it is useful to have a consistent framework to describe these two cases.
The thermodynamic cycle and notations of section~\ref{sec:restraints} are well-suited for that.

\section{Typical practical problem cases}

Below, we describe three typical situations that cover most of the binding cases where symmetry issues are encountered.
They include two ways in which binding can break symmetry (either a symmetric receptor or a symmetric ligand, the symmetry of which is lost in the complex), and a case where binding increases symmetry, when two identical molecules associate to form a symmetric homodimer.

\subsection{Case 1: symmetric receptor -- Equivalent sites}

\subsubsection{Equivalent binding sites on a protein complex}
\label{sec:insulin}

A common case of equivalent symmetric binding sites is encountered with protein multimers, when symmetric binding sites are found on the supramolecular complex. This is, for instance, the case of phenol binding to the insulin hexamer,\cite{Derewenda1989} which is relevant for pharmaceutical preparation of insulin, and has been recently studied by simulations.\cite{Palivec2017}

\begin{figure}[h!]
    \centering
    \includegraphics[width=0.5\textwidth]{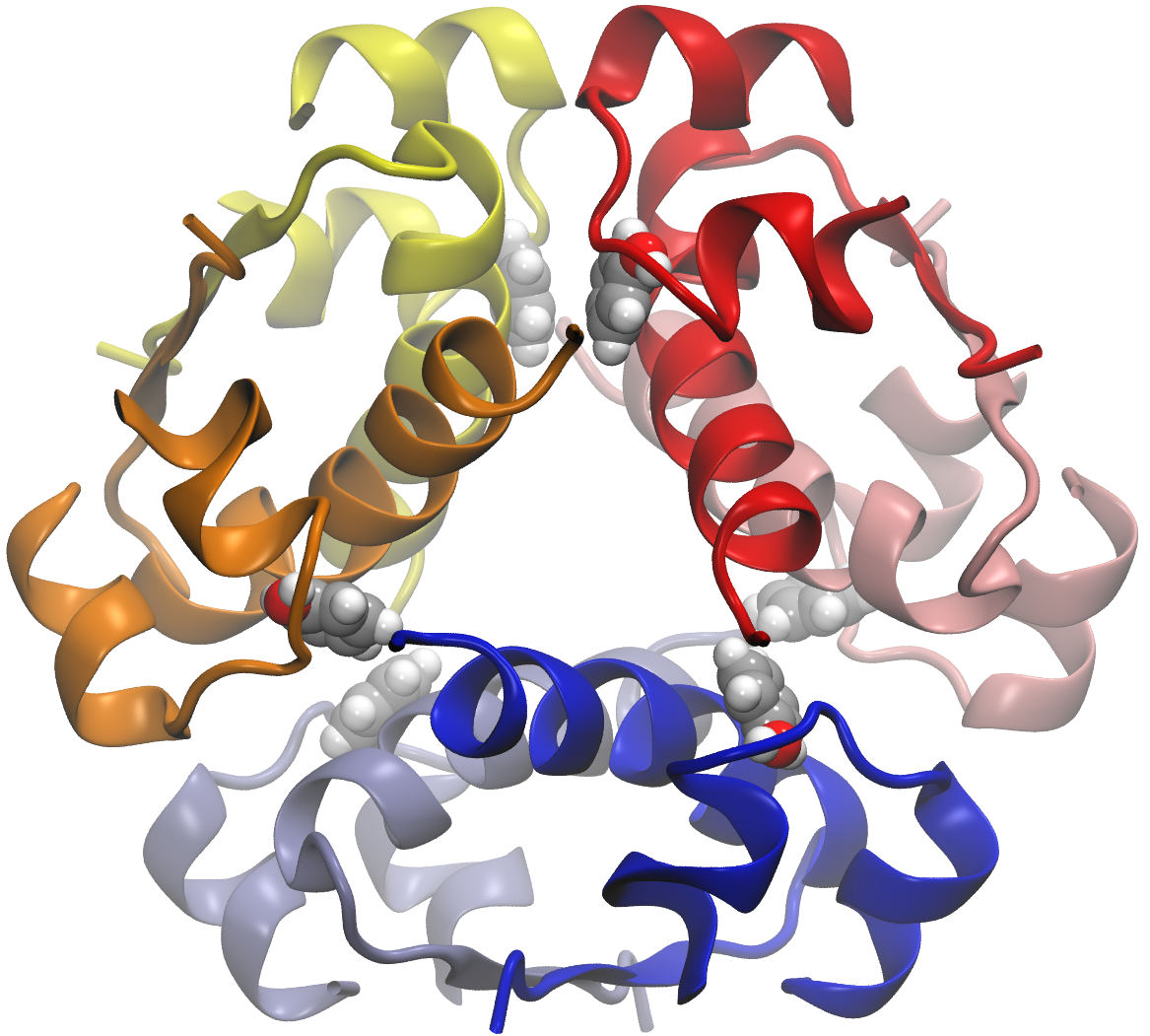}
    \caption{Structure of the insulin hexamer bound to six phenol molecules, rendered with VMD.\cite{Humphrey1996}}
    \label{fig:insulin}
\end{figure}

We will first consider the first binding free energy, that is, the binding of one ligand to the protein multimer.

Typically, in such a case, one would model the complex with a single ligand bound, and perform alchemical decoupling using, for instance, a flat-bottom or harmonic restraint on the distance between the ligand center-of-mass and that of selected protein atoms around the binding site. This restraint is asymmetric within the framework developed in section \ref{sec:sym_restraints}.

In practice, in the case of physically well-separated, symmetric binding sites, it is numerically easier to first consider binding to a single site, and evaluate the restraint contribution with respect to this single site definition, $\Delta G^{\text{rest}\rightarrow \text{1site}}_\text{coupled}$. This procedure yields the binding constant to a single site $K^{\text{1site}}$, which ignores the other, symmetric sites.
This is the \textit{microscopic} binding constant of the biophysics literature.
The overall first binding constant to any of of the $n$ identical sites is:\cite{Cantor1980} 
\begin{equation}
K_1 = n \; K^{\text{1site}} 
\label{eq:K1}
\end{equation}

The a posteriori ``symmetry correction'' included in Eq.~\ref{eq:K1} can be viewed in our framework as part of the restraint free energy contribution $\Delta G^{\text{rest}\rightarrow \text{site}}_\text{coupled}$:

\begin{equation}
    \Delta G^{\text{rest}\rightarrow \text{site}}_\text{coupled} = \Delta G^{\text{rest}\rightarrow \text{1site}}_\text{coupled} - RT\ln(n)
\end{equation}
where the term $-RT\ln(n)$ accounts for changing the site definition.

To predict binding of more ligands to the mono-liganded complex, one must then take into account the fact that this complex is less symmetric than the isolated receptor, so nothing more can be said without further simplifying assumptions. Typically, multimeric proteins like insulin are epected to exhibit cooperative binding.
If, however, one assumes independent binding in the $n$ different sites, then the binding constant of $i$ ligands to the receptor can be expressed as a general function of $K^{\text{1site}}$ as:\cite{Cantor1980}
\begin{equation}
K_i = \dfrac{n-i+1}{i} \; K^{\text{1site}} 
\end{equation}

In the more realistic framework of cooperative binding, one would have to simulate explicitly the second binding event, with sufficient timescales to allow relaxation of the interactions that couple the two sites, and taking into account further changes in symmetry, depending on which combination of sites is populated.
This is evidently a much more involved endeavor, and beyond our scope here.

\subsubsection{Cation binding to carboxylate groups: an example of a symmetric binding site and partial sampling.}
\label{sec:acetate}

In the case discussed above of symmetric binding sites on large objects such as proteins, sampling is usually attempted---with restraints chosen accordingly--- for only one of the binding sites at a time. However, symmetric ``receptors'' can also present two symmetric binding modes that are spatially close, in which case one would more spontaneously attempt sampling both modes at once. We will discuss in detail such a situation, showing how it relates and differs from the previous one, on the typical case of ion binding to a symmetric carboxylate group---acetate for the sake of simplicity---that we recently encountered, and for which we needed binding predictions to help interpret experiments.\cite{DeOliveira2020}

\begin{figure}[ht]
    \centering
    \includegraphics[width=0.45\textwidth]{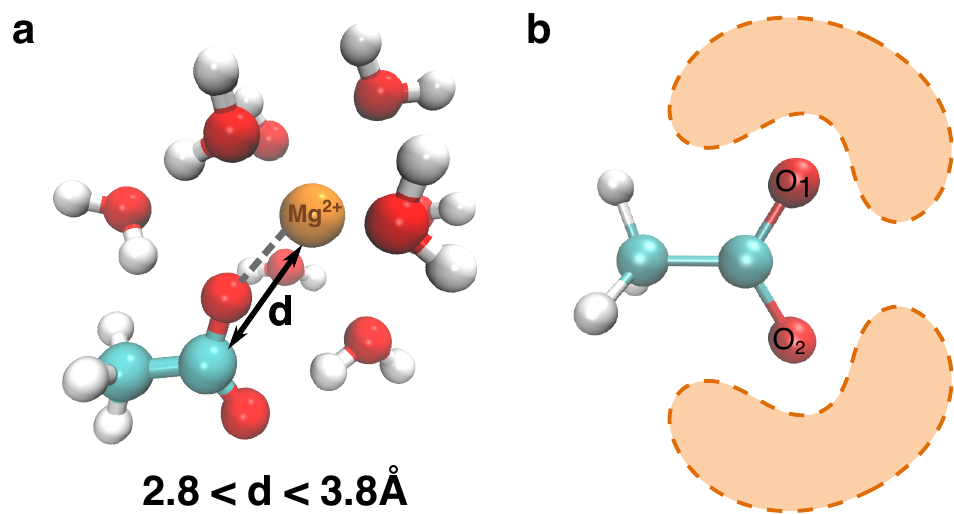}
    \caption{a) Simulation snapshot of a magnesium cation interacting with an acetate anion in a monodentate mode.  b) Qualitative scheme of the two symmetric monodentate binding lobes. The figures were prepared using the VMD visualization software.\cite{Humphrey1996}}
    \label{fig:acetatesym}
\end{figure}

Different modes of cation binding to carboxylate groups have been described.\cite{Dudev2007a,Einspahr1981} A cation can interact directly either with both carboxylate oxygen atoms in the bidentate binding mode, or with only one of the two oxygens in the monodentate mode. Binding can also occur through water molecules without any direct interactions, with the formation of solvent-shared ion pairs. For the purpose of this discussion, we now focus on the monodentate binding mode, our goal being to properly estimate the 1:1 monodentate binding of a cation (\textit{e.g.} Mg$^{2+}$) to acetate. The binding site corresponding to monodentate binding is symmetric, since the cation can interact symetrically with either of the two oxygen atoms as pictured in Fig.~\ref{fig:acetatesym}.

The distance $d$ between the carbon atom of the carboxylate group and the cation is the natural coordinate, previously used in the literature,\cite{Kumar2015e,DeOliveira2020,Daily2016c,Martinek2018} to define the monodentate binding mode and separate it from bidentate and solvent-shared binding modes. For instance, in the case of Mg$^{2+}$ binding to acetate, monodentate bound geometries correspond to $2.8<d<3.8$ \AA, bidentate complexes typically have $d<2.8$~\AA, and solvent-shared ion pairs correspond to $d>4$ \AA.

The binding free energy corresponding to the monodentate binding mode can then be calculated with the double decoupling method, using the thermodynamic cycle described in Fig.~\ref{fig:cycle}. Weak flat-bottom harmonic restraints acting only if $d>3.8$ \AA\ or $d<2.8$ \AA\ are imposed during the alchemical decoupling in the bound state to maintain the system in a monodentate geometry. The employed restraint is symmetric and does not restrict the sampling to only one of the two symmetric binding poses, unlike what was discussed previously for symmetric binding sites on protein multimers.

For this practical example, we use standard non-polarizable force fields both for acetate\cite{Martinek2018} and the magnesium ion.\cite{Li2013} The free energy associated with each alchemical transformation is obtained using the BAR\cite{Bennett1976} algorithm. Other simulation details are provided in section~\ref{sec:compdet}. Table~\ref{tab:DG} summarizes the free energies associated with each step of the thermodynamic cycle (Fig.~\ref{fig:cycle}), including the appropriate second order correction to $\Delta G^*_\text{bulk}$ due to the change in overall charge during the alchemical transformation.\cite{Simonson2016} $\Delta G^{\text{rest}\rightarrow \text{site}}_\text{coupled}$ is estimated on the fully coupled simulation window using Eq.~\ref{eq:DGrestbound}, the site being defined strictly as $2.8<d<3.8$~\AA.
Strictly speaking, this region contains the physically accessible site as well as regions that correspond to steric clashes and are effectively forbidden: it is therefore a suitable definition of the site for numerical purposes.
As expected with flat-bottom restraints that let the ligand evolve freely in the binding site, $\Delta G^{\text{rest}\rightarrow \text{site}}_\text{coupled}$ is small (in the present case, negligible). $\Delta G^{\text{V}^\circ\rightarrow\text{rest}} _\text{decoupled}$ is evaluated with Eqs.~\ref{eq:DGrest}-\ref{eq:DGrest_Q} by numerical integration of the restraining potential. The final standard binding free energy $\Delta G^\circ_\text{bind}=-37.2$ kJ~mol$^{-1}$ is then obtained combining all the different terms according to Eq.~\ref{eq:DGcycle}.

\begin{table*}[ht]
\begin{tabular}{|c|c|c|c|c|c|}
\hline
     Free energy term & $\Delta G^*_\text{bulk}$ & $\Delta G^*_{\mathrm{site}}$  & $\Delta G^{\text{V}^\circ\rightarrow \text{rest}}_\text{decoupled}$ & $\Delta G^{\text{rest}\rightarrow \text{site}}_\text{coupled}$ & $\Delta G^\circ_\text{bind}$  \\\hline
    kJ mol$^{-1}$ & 1726.7 (0.3) & 1769.7 (0.3) & 5.91 & 0.0 & $-$37.1 (0.6) \\\hline
\end{tabular}
\caption{Decomposition of the free energy terms of  Eq.~\ref{eq:DGcycle}, for the acetate-magnesium monodentate binding. }
\label{tab:DG}
\end{table*}

Since our definition of the ``bound state'' encompasses both of the two symmetric binding modes (to both oxygen atoms), the obtained binding free energy $\Delta G^\circ_\text{bind}$  directly measures the binding to both of the symmetric sites, \textbf{without need for any symmetry correction}, despite the symmetry of the binding site.

We did not comment so far on the actual conformations sampled during the alchemical simulation. In practice, in the coupled state and in the first windows of the alchemical transformation, only one of the two binding sites is visited, due to kinetic trapping of the cation. Intuitively, it would then be tempting to try and correct for this partial sampling, since only half of the bound configurations are visited in the fully coupled state. However, we argue that counter-intuitively and despite partial sampling, the decoupling free energy is not affected and no corrections are needed.
This derives from the fact that the BAR estimator that we employ depends only the potential energy distributions in each decoupling window. As discussed in \ref{sec:partial_sampling}, as long as the partial sampling does not modify the potential energy distribution, then the same free energy is obtained, irrespective of the sampling. We use the following numerical experiments to illustrate our point. \\
First, we reanalyze the data from our initial alchemical transformation (Table~\ref{tab:DG}) where only binding to O1 was sampled in the first few windows, and keep \textbf{in all windows} only the conformations corresponding to this binding mode. We now obtain $\Delta G^*_{\mathrm{site}}=1769.2 \pm 0.6 \text{ kJ mol}^{-1}$, which is not different, within the error bars of our calculations, from the value initially obtained using all the samples. 
To further illustrate our point, we can use a slightly different Mg$^{2+}$ force field,\cite{Callahan2010} so that both monodentate binding modes are sampled in all the windows of the alchemical transformation. We now evaluate $\Delta G^*_{\mathrm{site}}$ using, in all windows, only the conformations where the cation is closer to O1 [resp. O2]. The obtained $\Delta G^*_{\mathrm{site}}$ values do not significantly change---$1647.4 \pm 0.8$ and $1647.5 \pm 0.8$~kJ~mol$^{-1}$ respectively---and are identical within the error bars to that obtained using all the samples, $\Delta G^*_{\mathrm{site}}=1647.4 \pm 0.6$~kJ~mol$^{-1}$. 

In both these examples, we numerically verified that $\Delta G^*_{\mathrm{site}}$ does not depend at all on the extent of sampling of one or both symmetric binding modes in all or a few windows of the alchemical transformation. As detailed in section \ref{sec:partial_sampling}, this stems from the use of a free energy estimator that depends only on the distribution of comparison potential energies ($\Delta U_{AB}$) in each window. In our case, due to the symmetry of the two binding modes, the distribution is the same irrespective of the (partial) sampling, and thus the same is true for the computed binding free energy.

\subsection{Case 2: symmetric ligand -- Phenol binding to lysozyme.}
\label{sec:lyso}

In contrast with large biomolecules, small ligands (\textit{e.g.} phenol, benzene),  may exhibit global molecular symmetry. As an example, we consider the classic case of phenol binding to the engineered binding sites of the  lysozyme from phage T4.\cite{Eriksson1992, Merski2012} Various mutant lysozymes (L99A, L99A/M102Q, L99A/M102H) have been shown to bind a variety of organic molecules, such as phenol, with affinities that depend on the more or less polar nature of the cavity. This system has been extensively characterized experimentally and has become a benchmark for binding free energy (affinity) calculations.\cite{Hermans1997, Deng2006, Boyce2009,Mobley2017,Sakae2020}

\begin{figure}[ht]
    \centering
    \includegraphics[width=0.45\textwidth]{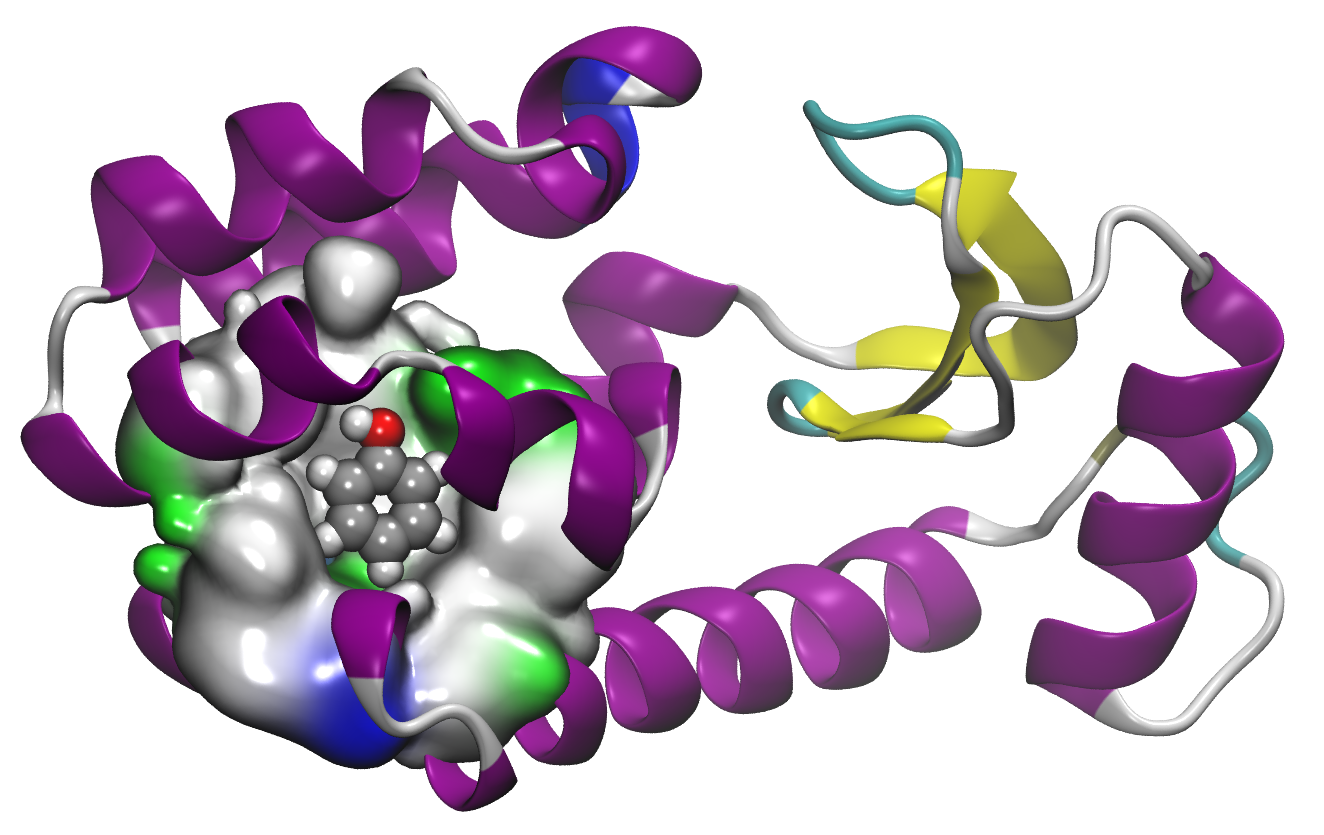}
    \caption{L99A/M102H mutant of T4 lysozyme complexed with phenol.
    The protein is rendered as cartoon representation and colored based on secondary structure elements. Residues from the binding site are rendered as a smooth surface and colored based on residue type (white: hydrophobic, green: polar, blue: basic).
    Phenol is rendered as spacefill and colored by element.
    Rendered using VMD\cite{Humphrey1996} based on PDB structure 4I7L.\cite{Merski2012}}
    \label{fig:lysophenol}
\end{figure}

Due to the symmetry of the phenol molecule, 180$^\circ$ rotation of the aromatic ring around the CO bond axis leads macroscopically to the exact same binding pose. However, the two symmetric binding poses can be artificially distinguished in simulations because of the labelling of each atom. 

In double decoupling free energy calculations, different kinds of restraints can be employed.
If simple harmonic (or flat-bottom) restraints on the distance between the phenol center of mass and the protein binding site are used, then the sampling of both symmetric poses is in principle (if not in practice) allowed. These restraints follow the symmetry of the bound state. In contrast, if orientational restraints (for instance, overall ligand RMSD with respect to the binding site\cite{Salari2018}) are employed, then the sampling is restricted to one of the two symmetric modes during the alchemical transformation. Such restraints are thus ``asymmetric'', in that they break the symmetry of the bound state.
As explained in Section \ref{sec:alchemy}, assuming numerical convergence, the different restraint schemes give the same value of $\Delta G^\circ_\text{bind}$, provided the restraint terms $\Delta G^{\text{rest}\rightarrow \text{site}}_\text{coupled}$ and $\Delta G^{\text{V}^\circ\rightarrow \text{rest}}_\text{decoupled}$ are properly taken into account.

However, we should underline that all protocols are not equivalent in terms of the rate of convergence of the $\Delta G^{\text{rest}\rightarrow \text{site}}_\text{coupled}$ term.
If the restraint is symmetric, then for the same reason as those developed in section \ref{sec:sym_restraints}, the restraint free energy is totally insensitive to partial sampling of one of the two modes, which makes its convergence dependent solely on relaxation within a binding pose, hence generally faster.
However, with an asymmetric restraint (allowing sampling of only one of the two modes), the estimation of the restraint free energy $\Delta G^{\text{rest}\rightarrow \text{site}}_\text{coupled}$ crucially depend on proper sampling of the two modes, because the perturbation now involves an asymmetric potential energy term, so comparison energies depend on the sampled mode (if the allowed mode is sampled in the restraint FEP, but not the forbidden mode, a spurious $RT \ln(2)$ appears in the free energy).
Hence, for numerical efficiency, \textit{it is generally advisable to use restraints of the same symmetry as the binding site definition}.
If, however, asymmetric restraints are preferred for any reason, and assuming that they forbid one of the poses entirely, then we recommend decomposing the evaluation of $\Delta G^{\text{rest}\rightarrow \text{site}}_\text{coupled}$ into two steps: first, estimate the restraint free energy with respect to only one of the binding poses; then add the analytical correction $-RT \ln(2)$ to take into account the existence of the other equivalent pose, a strategy similar to that adopted for symmetric binding sites in Section \ref{sec:insulin}.

\subsection{Case 3: symmetric complex -- Homodimerization}
\label{sec:homodimerization}

Biomolecules such as proteins frequently assemble into multimers, which are homomultimers (homodimers etc.) if they are formed by the assembly of identical molecules. 
The macroscopic thermodynamic description of homodimerization is slightly different from that of heterodimerization.
To pinpoint the effect of symmetry on the definition and calculation of a homodimerization constant, we describe a thought experiment wherein the very same physical process can be described as either homo- or heterodimerization.
We reason at the macroscopic level because it allows for a more direct connection with the macroscopic binding free energies that we seek to estimate.

Consider a solution containing solutes that can be arbitrarily considered of the same type $M$, or as distinct types $A$ and $B$---a concrete example would be methane molecules containing either $^{12}C$ or $^{13}C$, which could be considered identical or different depending on the method used to detect them.

We first consider the heterodimerization equilibrium:
$$A + B \rightleftharpoons AB$$
Suppose that the system is large enough that it obeys the law of mass action,\cite{Jong2011, JostLopez2020} and that the total concentration of both $A$ and $B$ is 1.
We also assume that $A$ and $B$ have exactly the same behavior with respect to dimerization.
The population of dimers can be thought of as evenly split between $AA$, $BB$, $AB$, and $BA$, the latter two being of course the same object; thus the heterodimer is twice as concentrated as each homodimer.
This can be arrived at by a symmetry argument: this is the distribution that would be obtained by random assignment of the labels $A$ and $B$ to otherwise identical molecules.
If we define  $x \equiv [AB]$, then $[A_2] = [B_2] = x/2$.
The free monomer concentrations are $[A] = [B] = 1 - 2x$.

We can write the law of mass action for heterodimerization:
\begin{equation}
K^\circ_{AB} \equiv \frac{[AB] C^\circ }{ [A] [B]} = \frac{x C^\circ }{ (1 - 2x)^2 }
\label{eq:kAB}
\end{equation}
If we compute a PMF $w_{AB}(r)$ between $A$ and $B$, we can directly apply Eq.~(\ref{eq:pmf}) to the heterodimerization, and obtain the value of 
$K^\circ_{AB}$.
Note that this requires no particular assumption on the possible homodimerization of species $A$ and $B$ separately.

Now we re-analyze the very same physical system, ignoring the labels  $A$ and $B$ and regarding all solutes as the generic monomer $M$.
The homodimerization equilibrium writes:
$$2M  \rightleftharpoons M_2 $$

And $[M_2] = [AB] + [A_2] + [B_2] = 2x$, and $[M] = [A] + [B] = 2(1 - 2x)$.

The law of mass action for homodimerization is:
\begin{equation}
K^\circ_{M_2} \equiv \frac{[M_2] C^\circ }{ [M]^2 } = \frac{2x C^\circ }{ [2(1 - 2x)]^2 }
\label{eq:kM2}
\end{equation}

Comparing equations \ref{eq:kAB} and \ref{eq:kM2}:
\begin{equation}
K^\circ_{M_2} = \frac{1}{2} K^\circ_{AB}
\end{equation}

That is, keeping the same interactions between the monomers, the homodimerization constant is half the heterodimerization constant.
This thought experiment may seem contrived, but any homodimerization equilibrium can be reduced to it by arbitrarily labeling each half of the monomers (isotope labeling is again a concrete analogy). Thus, this result is  valid for homodimerization in general.

Since $A$ and $B$ are identical to $M$ with respect to intermolecular interactions, the self-association PMF of $M$ is that of $A$ and $B$: 
$w_{MM}(r) = w_{AB}(r)$.
Therefore the homodimerization constant $K^\circ_{M_2}$ can be expressed by modifying Eq.~(\ref{eq:pmf}):
\begin{equation}
K^\circ_{M_2} = \frac{1}{2 V^\circ}\int_{\text{site}} 4\pi r^2 e^{-\beta w_{MM}(r)} \text{dr}
\label{eq:pmf_homodimerization}
\end{equation}

This expression has been arrived at by another route.\cite{Domanski2016}
However, on many occasions, the factor of 2 has been omitted, including by us.\cite{Henin2005, Henin2010}
The resulting discrepancy of $RT \ln(2)$ is small enough to be usually inconspicuous, given the error margins of computational---and experimental---estimates of binding free energies.

\section{Conclusion}

We have recalled the main exact theoretical results allowing for the practical determination of macroscopic, absolute binding free energies based on explicit-solvent molecular simulations.
Adding to the existing practical advice on how to perform absolute binding free calculations using alchemical transformations,\cite{Mey2020} we point to several controversial steps in the calculations that can lead to erroneous numerical treatment--even though the resulting error is often hidden in the large error bars of the alchemical transformation itself. We argue that a key step to obtain well-defined binding constants (or binding free energies) that can be compared to experimental data is to precisely define the binding site (or mode), the thermodynamics of which you wish to characterize. This should be done as consistently as possible with the experimental measurements that can be predicted by or compared with simulations. This often requires simplifications (such as assuming a two-state model), unless direct computation of the experimental signal is possible.

Another key point is to carefully correct for the restraints both in the decoupled and coupled state, in a way that is consistent with the site definition. 
We show that this clarifies several practical situations and, in the case of symmetric receptor or ligand, eliminates the need for a specific and often ill-interpreted symmetry correction term.
Instead, symmetry effects are accounted for when evaluating the restraint contributions to the free energy in both the coupled and decoupled states, and we discuss the potential pitfalls of different protocols.
We have also shown that counter to a very common intuition, in symmetric cases, estimators of the excess free energies of decoupling are not affected by total, partial, or non-existent sampling of some symmetry-equivalent modes.

Finally, we have shown that the computation of macroscopic constants for homodimerization equilibria requires a special correction factor, and explained its origin.

\section{Computational details}
\label{sec:compdet}
Computation of the binding free energy of Mg$^{2+}$ to acetate (\ref{sec:acetate}) was performed using the Gromacs 5.1.1 software.\cite{VanderSpoel2005} The simulation box contained one acetate anion, one cation and 1723 water molecules. The simulations were performed in the constant temperature/constant pressure (NpT) ensemble, using the same setup as previously reported.\cite{DeOliveira2020}
Water molecules were described by the SPC/E force field,\cite{Berendsen1987} and two different force fields\cite{Li2013,Callahan2010} for Mg$^{2+}$, giving rise to two different behaviors with respect to sampling of one or both symmetric sites, were used. Note that these force fields are known to strongly overestimate ion-acetate binding, as we discussed in a previous work,\cite{DeOliveira2020} and their use here is only meant to illustrate typical issues encountered in free energy calculations.
Free energy calculations were performed using the double decoupling procedure described in \ref{sec:alchemy}. The free energy difference associated with each alchemical transformation was reconstructed using the Bennett Acceptance Ratio\cite{Bennett1976} (BAR) method as implemented in Gromacs. The restraint used during the alchemical transformation in the bound state is a flat bottom well potential, flat for $2.8<d<3.8$, and harmonic on each side with a $k=100000$~kJ~mol$^{-1}$~nm$^{-2}$ force constant.

\begin{appendix}

\section{Theoretical connection between the dimer counting and PMF formalisms}
\label{sec:counting}

Here we derive the connection between Eq.~\ref{eq:pmf}, which expresses the binding constant in terms of an integral over the PMF, and dimerization statistics in a microscopic simulation system.

Most binding free energy calculations rely on simulations of a unitary system of volume $V$ containing 1 $R$ and 1 $L$.
Calling $x$ the probability of the bound state (or bound fraction) in that system, and naively applying the law of mass action (Eq.~\ref{eq:K_def}) would lead to:

\begin{equation}
K^\circ_\text{bind}=\frac{x}{(1-x)^2}\frac{V}{V^\circ} \;,
\label{eq:Kx1}
\end{equation}

where $V^\circ = 1/C^\circ$ is the standard volume.
However, it has been shown\cite{Jong2011} that this expression is not applicable to a microscopic simulation box containing a single R and L: the law of mass action results from statistics over many copies of R and L.
In a unitary system where the solutes do not interact outside of the bound state,\cite{JostLopez2020} $K^\circ_\text{bind}$ may instead be expressed as:

\begin{equation}
K^\circ_\text{bind}=\frac{x}{(1-x)}\frac{V}{V^\circ}
\label{eq:Kx2}
\end{equation}

We now show that this expression is closely related to the PMF-based expression of the binding constant (Eq.~\ref{eq:pmf}).
Consider the probability distribution $p(\vx)$ of the 3D ligand position with respect to the receptor, normalized to be 1 in the bulk. The framework of Eq.~\ref{eq:pmf} requires that the bound state be defined solely by the presence of $\vx$ in a region of space corresponding to the binding site, without regard to non-translational degrees of freedom.
We call $W$ the weight of the bound state in this distribution:
\begin{align}
W &= \int_\text{site} p(\vx) \text{d}\vx \\
&=  \int_\text{site} g(r) 4\pi r^2 \text{d}r
\end{align}
which we note is the integral in Eq.~\ref{eq:pmf}.
Under the assumptions of Eq.~\ref{eq:Kx2} that the partners are either bound or non-interacting ($w(r) = 0$ outside the site),\cite{JostLopez2020} the weight of the unbound state is:
\begin{equation}
\int_{\overline{\mathrm{site}}} p(\vx)d\vx = V - V_\text{site} \approx V \;, 
\end{equation}
with the common assumption that $V\gg V_\text{site}$, so that
\begin{equation}
\int_V p(\vx) \text{d}\vx = W + V
\end{equation}
Therefore
\begin{equation}
x=\frac{W}{W + V} \; \text{and} \; 1 - x=\frac{V}{W + V}
\end{equation}
and
\begin{equation}\frac{x}{1-x}=\frac{W}{V} \;,
\end{equation}
which, substituted into Eq.~\ref{eq:Kx2}, gives Eq.~\ref{eq:pmf}.

\begin{table*}[ht!]
\begin{tabular}{|c|c|c|c|c|c|}
\hline
Site def. & Restraint & $\Delta G^*_\text{bulk}-\Delta G^*_\mathrm{site}$  &
$\Delta G^{\text{V}^\circ\rightarrow \text{rest}}_\text{decoupled}$ &
$\Delta G^{\text{rest}\rightarrow \text{site}}_\text{coupled}$ & $\Delta G^\circ_\text{bind}$ \\\hline
asym & asym & $\Delta\Delta G^*$ & $-RT\ln(V_\text{rest}^1/V^\circ)$ & $0$ &
$\Delta\Delta G^* -RT\ln(V_\text{rest}^1/V^\circ)$\\
sym & asym & $\Delta\Delta G^*$ & $-RT\ln(V_\text{rest}^1/V^\circ)$ & $-RT\ln(n)$ & 
$\Delta\Delta G^* -RT\ln(V_\text{rest}^1/V^\circ)-RT\ln(n)$\\
sym & sym & $\Delta\Delta G^*$ &
$-RT\ln(n\times V_\text{rest}^1/V^\circ)$ & $0$ &
$\Delta\Delta G^* -RT\ln(V_\text{rest}^1/V^\circ)-RT\ln(n)$  \\
    \hline
\end{tabular}
\caption{\textbf{Decomposition of the absolute binding free energy (Eq.~\ref{eq:DGcycle}), in the case of binding modes with or without order-$n$ symmetry, with symmetric or asymmetric flat-bottom restraints.}
Simple analytical expressions apply because of assumptions listed in the text. In a more general case, restraint free energies may be estimated numerically, and the $RT\ln(n)$ terms listed here are, at convergence, accounted for by the numerical estimators, and need not be explicitly added. See section~\ref{sec:lyso} for a discussion of cases where these might not converge in practice, and strategies to handle it. }
\label{tab:rest_sym}
\end{table*}

\section{Analytical treatment of symmetry contributions to restraint free energies under simplifying assumptions}
\label{sec:rest_sym_analytical}

To make the symmetry contributions to the restraint free energies apparent, we describe a case where $\Delta G^{\text{V}^\circ\rightarrow \text{rest}}_\text{decoupled}$ and
$\Delta G^{\text{rest}\rightarrow \text{site}}_\text{coupled}$ admit simple analytical expressions.
In most practical applications, at least one of these terms has to be computed numerically.
Suppose that we want to estimate the binding affinity of a bound state that comprises either $n$ symmetry-equivalent binding poses (symmetric ligand) or $n$ symmetry-equivalent binding sites (symmetric receptor): these two cases are largely equivalent from a formal perspective.
Crucially, we apply flat-bottom restraints that do not perturb binding to an individual mode.
Different definitions of both the site and the restraints are possible. Table \ref{tab:rest_sym} summarizes the value of the different free energy terms depending on these choices. As stated above, the values of the excess free energies of decoupling, $\Delta G^*_\text{bulk}$ and $\Delta G^*_\mathrm{site}$, are independent from the symmetry of the restraint scheme or that of the site definition. We call $V_\text{rest}^1$ the restraint volume for one mode.
If we use an asymmetric site definition, \textit{i.e.} limited to a single binding site or binding pose, then we naturally employ asymmetric restraints (first line of Table \ref{tab:rest_sym}) and the obtained binding free energy is then:
\begin{equation}
\Delta G^\circ_\text{bind, asym} = \Delta\Delta G^* -RT\ln(V_\text{rest}^1/V^\circ)\;.
\end{equation}

In contrast, if we define the binding site as symmetric, then different restraints can be used (last two lines of Table~\ref{tab:rest_sym}), either obeying or breaking the symmetry of the binding site. Both setups eventually yield the same standard free energy of forming the $n$-fold-symmetric complex:
\begin{equation}
\Delta G^\circ_\text{bind} = \Delta\Delta G^* -RT\ln(V_\text{rest}^1/V^\circ)-RT\ln(n)\;.
\end{equation}
\end{appendix}

\section*{Acknowledgments}
This work was supported by the ‘‘Initiative d’Excellence’’ program from the French State (Grants ‘‘DYNAMO’’, ANR-11-LABX-0011, and ‘‘CACSICE’’, ANR-11-EQPX-0008). Computational work was performed using HPC resources from LBT/HPC. We thank our colleagues H. Martinez-Seara, P. Jungwirth and V. Palivec  (UOCHB, Prague, CZ), C. H. Robert (LBT, IBPC, Paris), and G. Brannigan (Rutgers Camden, NJ, USA) for stimulating discussions.

\section*{Data Availability Statement}
The data that support the findings of this study are openly available in Zenodo at \\ http://doi.org/10.5281/zenodo.4519498.

\bibliography{free_energy}

\end{document}